\documentclass[12pt]{article}
\usepackage{amsmath,amssymb,multirow,latexsym,amsfonts,graphicx}
\usepackage{color}
\usepackage{cite}
\usepackage{slashed}
\usepackage{comment}
\usepackage{multicol}
\newcommand{\be}{\begin{equation}}
\newcommand{\ee}{\end{equation}}
\newcommand{\bea}{\begin{eqnarray}}
\newcommand{\eea}{\end{eqnarray}}

\newcommand{\nc}{\newcommand}
\nc{\vect}{\mathbf}

\def\P{Poincar\'e }
\def\Sch{Schr\"odinger }

\begin{document}
\renewcommand {\theequation}{\thesection.\arabic{equation}}
\renewcommand {\thefootnote}{\fnsymbol{footnote}}
\vskip1cm
\begin{flushright}
\end{flushright}
\vskip1cm
\begin{center}
{\Large\bf Noncommutative Time in Quantum Field Theory\\\vskip0.3cm }

\vskip .7cm

{\bf{\large{Tapio Salminen and Anca Tureanu}}\\

\vskip .7cm

{\it\large Department of Physics, University of Helsinki,\\ P.O. Box
64, FIN-00014 Helsinki, Finland
}}

\end{center}

\vskip1cm
\begin{abstract}
We analyze, starting from first principles, the quantization of field theories,
in order to find out to which problems a noncommutative
time would possibly lead. We examine the problem in the interaction
picture (Tomonaga-Schwinger equation), the Heisenberg picture (Yang-Feldman-K\"all\'{e}n equation) and the path integral approach. They
all indicate inconsistency when time is taken as a noncommutative
coordinate. The causality issue appears as the key aspect, while the unitarity problem is subsidiary.
These results are consistent with
string theory, which does not admit a time-space noncommutative
quantum field theory as its low-energy limit, with the exception of light-like
noncommutativity.
\end{abstract}

\vskip1cm


\section{Introduction}
\setcounter{equation}{0} The role of time has varied dramatically in
the history of physics. In pre-relativistic physics it had a special
role and was disconnected from spatial directions. In special
relativity the notions of space and time are combined to form
Minkowski space and in the recent attempts to construct theories of quantum
gravity it again seems that time becomes truly special. In the
extreme, time might be demoted to a mere parameter in low energy
theory, deprived of any physical meaning \cite{DeWitt:1967yk,
Barbour:2000ad}.

In quantum field theories the implications of Lorentz invariance are
considered essential properties of nature. However, with
the development of quantum field theories on noncommutative
space-time \cite{DFR1, DFR2,SW} we are forced to abandon our preconceived
notions of Lorentz invariance and the role of time needs to be re-evaluated.
The Weyl-Moyal space-time which most often replaces Minkowski
space in noncommutative theories is defined through the
Heisenberg-like commutator:
\be [\hat x^\mu ,\hat x^\nu]=i\theta^{\mu \nu}\\, \ee
where $\theta$ is a constant antisymmetric matrix (not a tensor).
Taking $\theta$ constant has its disadvantages, most notably the
evident loss of Lorentz invariance. These theories exhibit other, so-called {\it twisted} symmetries, most importantly the twisted \P
invariance \cite{CKNT,CPrT} that allows us to retain some of the
essential features usually associated with Lorentz invariance,
such as irreducible particle representation, spin-statistics and the
CPT theorems \cite{CKNT}-\cite{CKTZZ}. Albeit very useful, these
symmetries are not further elaborated in this paper (for further
discussion we refer the reader to a recent work and references therein
\cite{Chaichian:2008ge}).

In this paper it is our purpose to consider Lorentz-covariant theories
with noncommutative time, of the Doplicher-Fredenhagen-Roberts type
\cite{DFR1,DFR2}. These works had a seminal effect in the study of
space-time noncommutativity and it should be pointed out that the
noncommutativity of time has an essential role in their
motivation.

We use a tensor $\theta$-matrix in four dimensions. Without any loss of
generality, in a given frame, it can be written in the form
\begin{eqnarray}\label{theta}
\theta^{\mu\nu}=\left(
\begin{array}{cccc}
0 &\theta & 0  & 0 \\
-\theta & 0 & 0  & 0 \\
0 & 0  &0 & \theta '\\
0 & 0  & -\theta' & 0
\end{array}
\right).
\end{eqnarray}
In field theories constructed in noncommutative space-time the usual products of functions need to be replaced by the $\star$-product:
\be\label{int rep}(f\star g)(x)=\int d^4 y\ d^4 z\ \mathcal K(x;y,z)
f(y) g(z)\,,\ee
where
\be \label{kernel}\mathcal
K(x;y,z)=\frac{1}{\pi^{4}\det\theta}\exp[-2i(x\theta^{-1}y+y\theta^{-1}z+z\theta^{-1}x)]\,,\ee
with
the notation
$x\theta^{-1}y=x^\mu(\theta^{-1})_{\mu\nu}y^\nu$. The integral representation can also be expressed in a form which is insensitive to the singularity of the $\theta$-matrix, as follows:
\be
(f\star g)(x)=\frac{1}{(2\pi)^4}\int d^4 y\ d^4 z\ f\left(x-\frac{1}{2}\theta y\right)g\left(x+z\right)e^{-iyz}\,,
\ee
with the obvious notation $(\theta y)^\mu=\theta^{\mu\nu}y_\nu$.
Differing slightly
in its support properties \cite{JGB,test_func}, but sometimes simpler to use, is the differential form of the
$\star$-product:
\be\label{starprod} \left(f \star g\right)(x) \equiv
\left[f(x)e^{\frac{i}{2}\theta_{\mu \nu}
\frac{\overleftarrow\partial}{\partial x _\mu}
\frac{\overrightarrow\partial}{\partial y_\nu}} g(y)\right]_{x=y}\,.
\ee
We shall mostly use the integral
representation \eqref{int rep}, \eqref{kernel}, but we shall refer as well to the differential
form \eqref{starprod}. Though the two representations are not fully equivalent,
the latter being only an asymptotic expansion of the first, it turns
out that the features which we shall explore are not sensitive to
the choice of the representation of the $\star$-product, but rather
to the nonlocality in time.

The purpose of this paper is to clarify the issues differentiating
the theories where the time coordinate is taken to be either
commutative or noncommutative. There has been much debate on this
issue and the view on whether theories with noncommutative
time can be consistently defined has shifted frequently.

The main problems connected with the noncommutativity
of time are causality and unitarity. There is consensus in the literature that causality is violated both
macroscopically  \cite{SST} and microscopically (see, e.g., \cite{LAG}). However, concerning unitarity, though shown to be violated
perturbatively in \cite{GM} in the path integral formulation (using what is sometimes called ``na\"ive Feynman rules''), there
have been various attempts to restore it in operator formulation,
both in the interaction picture \cite{Bahns2,LS02,Grosse03} and in the
Heisenberg picture (Yang-Feldman-K\"all\'en formalism) \cite{Doescher}. However, it appears that out of the two problems the first one is the key-issue, while the latter is one manifestation of the first. Fixing the latter problem does not automatically rid us of other contradictions.

It has been very puzzling that some of the approaches in defining a
unitary S-matrix, apparently consistent, are intimately
connected to other approaches, which are manifestly inconsistent.
The best examples here are the Heisenberg picture \cite{Bahns2,BahnsThesis} and the path
integral formulation \cite{Fuji}. Also, within the interaction picture has
appeared the bizarre aspect that the equation of motion of the
fields is not the same as in the ordinary QFT \cite{BahnsThesis}. As a peculiar technical aspect, it should be mentioned that the Euclidean and the Minkowski versions seem to be quite different when time is not commutative, since there exists no Wick rotation to relate them.

The use of quasiplanar Wick products proposed in \cite{Bahns3}, although ingenious when considering the locality of the interacting theory, leads to nonlocality in the asymptotic \emph{in} and \emph{out} fields. In a way, the problem is shifted from the interacting theory to the free theory, but again full consistency is not achieved and the theory thus constructed has very peculiar dispersion relations, involving such features as negative group velocities and momentum-dependence of the mass of the asymptotic fields \cite{Bahns3}.

In \cite{LS02} the time-ordered perturbation theory (TOPT) was introduced for theories with noncommutative time, concluding that the construction leads to a unitary theory. In \cite{Ohl} it was shown that the noncommutative QED thus constructed does not satisfy Ward identities, therefore this procedure for recovering unitarity is not consistent either.

The issue of unitarity in TOPT formulation was taken up also in \cite{Grosse03} and the problem of causality was revisited. The conclusion was that apparently causality and unitarity are mutually exclusive when time is noncommutative. Remark, however, that even if certain constructions may appear to be unitary, causality is still violated -- the use of causal Feynman propagators in the path integral approach, for example, does not render the theory causal, as can be seen from the macrocausality analysis of \cite{SST}.

The string theoretical analysis \cite{SST2} shows, however, that for a particular type of time-space noncommutativity, i.e. the light-like noncommutativity, the low-energy limit as noncommutative quantum field theory exists and is free of the unitarity problem \cite{AGM}. The quantization of the theory as a field theory has been proven to be possible \cite{NCllUVIR} by making use of the light-cone coordinates. The distinguishing feature of light-like noncommutativity is that the {\it light-cone time} can be chosen as commutative and then the quantization of the theory resembles the quantization of a space-space noncommutative theory in light-cone coordinates. The notion of physical causality remains still under question. It is also interesting that, while for ordinary relativistic theories the covariant quantization and the light-front frame quantization lead to the same result, in the case of light-like noncommutativity the covariant quantization is impossible, but the quantization in light-cone coordinates (with a nonrelativistic flavour to it) can be achieved and the results are in agreement with the string theory analysis.

The noncommutative quantum field theory is a special case of the {\it nonlocal quantum field theories}, which were thoroughly studied especially in the 1950s, in the context of strong (meson theory) interactions.
H. Yukawa proposed a nonlocal field theory \cite{Yukawa}, with the hope that the divergences in the meson theory could be better controlled in a relativistic theory that takes
explicitly into account the finite extension of the constituent particles. In a subsequent paper C. Bloch \cite{Bloch} made clear that there is no hope of putting such a theory into Hamiltonian form, i.e. to find a corresponding Schr\"odinger equation (actually, the Tomonaga-Schwinger equation in the interaction picture), but instead he proposed to use the new at the time Yang-Feldman-K\"all\'en formulation \cite{Yang-Feldman,Kallen} in the Heisenberg picture. Shortly afterwards, in a paper by P. Kristensen and C. M{\o}ller \cite{Kris}, an allegedly convergent meson theory was put forward. This work was the starting point for an intensive study of nonlocal quantum field theories \cite{Hayashi}, which after a few years indicated that the Yang-Feldman-K\"all\'en approach in this case appeared also to be flawed. A revival of the subject came with the works of R. Marnelius in the beginning of the 1970s \cite{Marnelius_action, Marnelius_Smatrix}, on the (im)possibility of defining an $S$-matrix in quantum field theories with nonlocal interactions.

In this paper we return to first principles and discuss relativistic quantum field theories with noncommutative time in the interaction picture, the Heisenberg picture and in the path integral formulation. We shall highlight mainly the inconsistencies and connect them in the different pictures.
%

In Sec. II we consider the difficulties encountered in the interaction picture, where the microcausality condition (in terms of Hamiltonians of interaction) is also the integrability condition for the equations of motion (Tomonaga-Schwinger). Since causality is violated, it turns out that the equation of motion has non-unique solution. We find also that energy-momentum is not conserved, in spite of the translational invariance of the star-product. We also comment on the inapplicability of Matthew's theorem when the number of time derivatives is infinite.
In Sec. III we discuss the unitarity problem in the Heisenberg picture, as a special case of the relativistic nonlocal field theories of Ref. \cite{Marnelius_Smatrix}. The conclusion of Ref. \cite{Marnelius_Smatrix}, that a unitary $S$-matrix cannot be defined in this context, applies also to our case. The deviations appear in the fourth order in the coupling constant, and they surface either as nonunitary $S$-matrix, or as energy-momentum nonconservation. In Sec. IV we review the problems arising in the path integral approach \cite{Fuji}, which lacks unitarity and from which it is impossible to derive any canonical structure. The pervasive feature leading to all these dead ends is the lack of causality, which appears to be the key-problem in constructing a consistent Lorentz invariant quantum field theory in noncommutative space-time.

\section{Noncommutative time and the interaction picture}
\setcounter{equation}{0}

The defining aspect of the interaction picture is that the
total Hamiltonian of an interacting system can be split into the free
part and the interaction part, such that the operators evolve by the
free Hamiltonian according to the Heisenberg equation of motion,
while the states evolve by the interaction Hamiltonian, according
to the Schr\"odinger equation
\be\label{Schr_eq} i\frac{\partial}{\partial t} \Psi(t) = \int
d^3x\,\mathcal H_{int}(x)\Psi(t)\,,\ \ \ \ x=(t,\mathbf x)\,. \ee
Being nonrelativistic, the \Sch equation singles out the time and
the approach is not manifestly Lorentz covariant. Moreover, the
equation is nonlinear and it can not in general be solved to check the
covariance. The manifestly covariant formulation was proposed by
Tomonaga \cite{tomonaga} as a super-multiple-time theory, inspired
by Dirac's multiple-time approach in quantum mechanics. The
Tomonaga-Schwinger equation \cite{schwinger,tomonaga}  (see also
\cite{Nishijima_book,schweber}) is the basis of this formulation and
various issues which are not deducible in the single-time theory are
made transparent in the covariant approach.

Since we are interested in a Lorentz-covariant formulation of
noncommutative quantum field theory, it is natural to start from
Tomonaga-Schwinger, rather than from \Sch equation.

\subsection{Solutions of Tomonaga-Schwinger equation}

In ordinary QFT, the idea behind the covariant
quantization in the interaction picture is to assign an individual time
coordinate to each point in space, such that the set of points
$t_{xyz}=t$ define a three-dimensional hypersurface $\sigma$. Thus,
the single-time \Sch equation \eqref{Schr_eq} is split into an infinite
set of equations, known under the name of Tomonaga-Schwinger
equation, which reads, with the above notation:
\be\label{t-s} i\frac{\delta}{\delta \sigma(x)} \Psi[\sigma] =
\mathcal H_{int}(x)\Psi[\sigma]\,, \ee
with the boundary conditions
\be\Psi[\sigma_0]=\Psi\,.\label{TS_boundary_cond}\ee
Then Eq. \eqref{t-s} with the boundary condition
\eqref{TS_boundary_cond} represent a well-posed Cauchy problem.
 The existence of a {\it unique} solution
for the Tomonaga-Schwinger equation is ensured if the integrability
condition
\be\label{integr} \frac{\delta^2\Psi[\sigma]}{\delta
\sigma(x)\delta\sigma(x')}-\frac{\delta^2\Psi[\sigma]}{\delta
\sigma(x')\delta\sigma(x)}  = 0, \ee
with $x$ and $x'$ on the surface $\sigma$, is satisfied. This
integrability condition \eqref{integr}, plugged into \eqref{t-s}, implies
\be\label{inv com rule}[\mathcal H_{int}(x),\mathcal
H_{int}(x')]=0.\ee
Since in the interaction picture the field operators
satisfy free-field equations, they automatically satisfy Lorentz invariant
commutation rules. The Lorentz invariant commutation relations are
such that (\ref{inv com rule}) is fulfilled only when $x$ and $x'$
are space-like separated,
\be(x-y)^2<0\,,\ee
 i.e. when $\sigma$ is a space-like surface. As
a result, {\it the integrability condition \eqref{inv com rule} is
equivalent to the microcausality condition for local relativistic
QFT}. When the surfaces $\sigma$ are hyperplanes of constant time,
the Tomonaga-Schwinger equations reduce to the single-time \Sch
equation \eqref{Schr_eq}.

Having the integrability condition fulfilled, one can write the
formal solution
\be \Psi[\sigma]=U[\sigma,\sigma_0]\Psi\,, \ee
with the unitary  operator $U[\sigma,\sigma_0]$ satisfying
\be\label{t-s_evolution_op} i\frac{\delta}{\delta \sigma(x)}
U[\sigma,\sigma_0] = \mathcal H_{int}(x)U[\sigma,\sigma_0]\,,\  \ \
\ \ U[\sigma_0,\sigma_0]=1.\ee
The covariant integration of the Tomonaga-Schwinger equation
\eqref{t-s_evolution_op} leads to the $S$-matrix in the interaction
picture:
\be
S=U[\infty,-\infty]=1+\sum_{n=1}^\infty\frac{(-i)^n}{n!}\int_{-\infty}^\infty\ldots
\int_{-\infty}^\infty\, d^4x_1\ldots d^4 x_n\, T[\mathcal
H_{int}(x_1)\ldots \mathcal H_{int}(x_n)]\,,\ee
with the Hamiltonian densities time-ordered under the integral
by the operator $T$.

The space spanned by the solutions of the Tomonaga-Schwinger
equation in different noncommutative spaces was considered in
\cite{Chaichian:2008ge}. Here we briefly revisit the results for
theories with noncommutative time, i.e. with the $\theta$-matrix in the form
\eqref{theta}. In the noncommutative case, the use of the
interaction picture has the advantage that the free-field equations
for the noncommutative fields are identical to the
corresponding free-field equations of the ordinary case. Here our
interest lies in what replaces the space-like surface $\sigma$ in
Weyl-Moyal space-time, i.e. for which type of separation of the coordinates $x$ and $y$ the integrability condition
is satisfied. The Tomonaga-Schwinger equation in
the noncommutative case reads:
\be\label{NCTS} i\frac{\delta}{\delta \sigma'} \Psi[\sigma' ] =
\mathcal H_{int}(x)_\star\Psi[ \sigma']\,, \ee
where $\sigma'$ is to be determined and we make a simple choice for $\mathcal H_{int}(x)_\star$ as
\be
 \mathcal H_{int}(x)_\star = \lambda[\phi(x)]_\star^n
= \lambda\phi(x) \star \phi(x) \star \ldots \star \phi(x)
\,.\label{Hamiltonian} \ee
The fields $\phi(x)$ satisfy
free-field equations and the Hamiltonian of interaction is built up
by inserting $\star$-products in between the fields. Then the integrability condition
for \eqref{NCTS} turns out to be
\be\label{int} \left[ \mathcal H_{int}(x)_\star,\mathcal
H_{int}(y)_\star \right] = 0\,, \quad \text{for} \,\, x,y\in \sigma'
\,. \ee
Using the integral representation of the $\star$-product we can write \eqref{int} as
\begin{eqnarray}\label{NC integr cond}
\lambda^2\bigl[(\phi\star \ldots \star \phi)(x), (\phi \star
\ldots \star \phi)(y) \bigr]&=&\lambda^2\int \prod_{i=1}^n
d a_i \,\mathcal K( x; a_1,\cdots, a_n) \int
\prod_{j=1}^n d b_j \, \mathcal K( y;
b_1,\cdots, b_n) \cr
&\times&\bigl[\phi(a_1)\ldots \phi(a_n), \phi(b_1)\ldots \phi(b_n) \bigr]\,.
\end{eqnarray}
Furthermore, the commutator of products of fields appearing in (\ref{NC integr
cond}) is written as the sum of products of fields at various
space-time points multiplied by a commutator of fields. A
typical term contains products of the type:
\be\label{factor} \phi(a_1)\ldots \phi(a_{n-1})\phi(b_1)\ldots
\phi(b_{n-1})\bigl[\phi(a_n), \phi(b_n) \bigr]\,.\ee
The fields at each different point are independent, since they are systems
with an infinite number of degrees of freedom. As a result, their
products will also be independent. Eq. (\ref{NC integr cond})
becomes a sum of independent products of fields, whose coefficients
have to vanish identically in order for the whole sum to vanish.
Since the kernel can not vanish, the necessary condition is
for the commutators of fields to be zero at every point,
\be \bigl[\phi(a_i), \phi( b_j) \bigr] =\Delta(a_i-b_j)=0 \,, \label{cond_a_b}\ee
since in the interaction picture the field $\phi$ satisfies the same
free-field equations and the invariant commutation relations as in
the ordinary case. Obviously, $\Delta(a_i-b_j)$ is the causal $\Delta$-function of ordinary QFT. The condition \eqref{cond_a_b} is satisfied outside of the
mutual light-cone
\be\label{light-cone cond} ({a_i^0}-{b_j^0} )^2-({a_i^1}-{b_j^1} )^2
- ( {a_i^2}-{b_j^2})^2- ({a_i^3}-{b_j^3} )^2 < 0 \,. \ee
In order to satisfy \eqref{NC integr cond}, it is necessary that
\eqref{light-cone cond} holds for all values of ${{a_i}^k}$
and ${{b_j}^l}$. However, since the coordinates are integration
variables in the range
\be 0\leq ( {a_i^k}-{b_j^k} )^2 < \infty\,, \ee
the requirement \eqref{light-cone cond} is clearly not satisfied for the whole space of $a_i^k$ and $b_j^k$. This in turn means that the integrability condition \eqref{NC integr cond} is not satisfied for any $x$ and $y$.
Thus, the Tomonaga-Schwinger
equation does not have a uniquely determined solution in the case of
time-space noncommutative quantum field theory.

The fact that the condition \eqref{int} is not satisfied in general is a special case of the fact that the commutator of ``local'' observables composed with the $\star$-product, $\left[ \mathcal O_\star(x),\mathcal
O_\star(y) \right]$, does not vanish for any $x$ and $y$. This is the microscopic manifestation of the violation of causality \cite{CNT}, appearing at the macroscopic level \cite{SST}.

\subsection{The time evolution of the Hamiltonian of interaction: Nonconservation of energy}\label{energy_noncons}

The violation of microcausality has a peculiar effect also on the time evolution of the Hamiltonian of interaction. In the interaction picture, the time-dependence of the Hamiltonian of interaction of a conservative system is an artefact of the picture itself. The time evolution of any operator is given by:
\be\label{op_t_evol} i\frac{dA^I(t)}{dt}=[A^I(t), H_0]+iU_0^\dagger\frac{\partial A^S}{\partial t}U_0, \ \ \ \ U_0=e^{-iH_0t},\ee
where $A^I= U_0^\dagger A^S U_0$ is the operator in the interaction picture, $A^S$ is the same operator in the Schr\"odinger picture and $H_0$ is the free Hamiltonian.

Let us take
the simplest example of a Hermitian scalar field with $\lambda\phi^3$-interaction:
\be\label{time_evol_int_p} \mathcal H(x) = \frac{1}{2}\pi^2(x)+
\frac{1}{2}(\partial_i \phi(x))^2+ \frac{1}{2}
m^2\phi^2(x)+\frac{\lambda}{3!}\phi^3 (x)\equiv \mathcal H_0(x)+
\mathcal H_{int}(x)\,, \ee
where $\pi(x)= \frac{\partial
\mathcal L}{\partial(\partial_0 \phi(x))}=\partial_0\phi(x)$ and consider the time evolution of $ H_{int}(t)$. Since the system is conservative, $H^S_{int}$ is obviously time-independent, and thus the second term in the r.h.s. of \eqref{op_t_evol} drops out. By performing a direct calculation, using the usual equal-time commutation relations, we obtain
\begin{align}
\left[  H_{int}(t), H_0(t) \right] \notag &= \frac{1}{2}\int d^3x\; d^3y \,\left[\frac{\lambda}{3!}\phi^3(t,\vect x),\pi^2(t,\vect y) +(\partial_i \phi(t, \vect y))^2 + m^2\phi^2(t,\vect y)\right] \notag \\
&=\frac{\lambda}{2\cdot 3!} \int d^3x\; d^3y\, \left[\phi^3(t,\vect x),\pi^2(t,\vect y)\right] \notag \\
&=\frac{i\lambda}{2\cdot 3!} \int d^3x \; 2\left( \pi(t,\vect x)  \phi^2(t,\vect x) + \phi(t,\vect x)\pi(t,\vect x)\phi(t,\vect x)+ \phi^2(t,\vect x) \pi(t,\vect x)\right)\notag\\
&= \frac{i\lambda}{3!} \int d^3x \; \left( \partial_0\phi(t,\vect x)  \phi^2(t,\vect x) + \phi(t,\vect x)\partial_0\phi(t,\vect x)\phi(t,\vect x)+ \phi^2(t,\vect x) \partial_0\phi(t,\vect x)\right)\notag\\
&=i \frac{d}{dt}   H_{int}(t) \,,
\end{align}
which is an expected tautology.

In the case of noncommutative quantum field theory, but with commutative time, the $\star$-product does not contain time-derivatives, therefore there is no ambiguity in defining the momentum conjugated to the field as $\pi(x)= \frac{\partial
\mathcal L_\star}{\partial(\partial_0 \phi(x))} $, through which we switch from Lagrangian to Hamiltonian formalism. We recover the expected result:
\begin{align}
\left[  H_{int}^\star(t), H_0(t) \right] &= \frac{\lambda}{3!}\int d^3 x d^3y \,\left[ (\phi\star\phi\star\phi)(t, \vect y), \mathcal H_0(t, \vect x)\right]\notag\\& =\frac{\lambda}{3!} \int d^3x\,d^3y \,\left[ (\phi\star\phi\star\phi)(t, \vect y), \frac{1}{2}\pi^2(t, \vect x)\right] \notag\\
= \frac{i\lambda}{3!}&\int d^3x\, d^3y \int d^3 a_1 \,d^3 a_2\, d^3 a_3 \frac{1}{\pi^4\det \theta} \mathcal K( y; a_1,a_2,a_3) \notag\\
& \times\big( \delta(\vect x - \vect a_1) \pi(x)\phi(a_2)\phi(a_3) + \delta(\vect x - \vect a_2) \phi(a_1)\pi(x)\phi(a_3) + \delta(\vect x - \vect a_3) \phi(a_1)\phi(a_2)\pi(x)\big) \notag\\
=\frac{i\lambda}{3!}& \int  d^3y \, \left( \left( \pi\star \phi\star \phi\right)  (y) +
\left( \phi\star \pi\star \phi\right) (y) +\left(  \phi\star \phi\star \pi\right) (y)
\right)=i\frac{d}{dt} H_{int}^\star\,.
\end{align}

In the case of noncommutative time, there are
problems from the beginning in defining the conjugate momentum $\pi(x)= \frac{\partial
\mathcal L_\star}{\partial(\partial_0 \phi(x))} $, as the total Lagrangian density
contains infinitely many time derivatives. Therefore,  the Lagrangian density
\be\label{total_Lagr_int_p} \mathcal L_\star(x) = \frac{1}{2}\partial^\mu\phi(x)\partial_\mu\phi(x)-
 \frac{1}{2}
m^2\phi^2(x)-\frac{\lambda}{3!}\phi^3_\star (x)\ee
and the Hamiltonian density
\be\label{total_Hamilt_int_p} \mathcal H_\star(x) = \frac{1}{2}\pi^2(x)+
\frac{1}{2}(\partial_i \phi(x))^2+ \frac{1}{2}
m^2\phi^2(x)+\frac{\lambda}{3!}\phi^3_\star (x)\,, \ \ \ \pi(x)=\partial_0\phi(x)\ee
do not describe the same dynamics \footnote{In the context of constrained systems, in Ref. \cite{GKL} the Hamiltonian formulation of the dynamics described by the Lagrangian \eqref{total_Lagr_int_p}, essentially following the prescription of \cite{LV}, was derived. Also, the usual ``na\"ive'' Feynman rules are recovered in the path integral quantization with constraints.}. We have omitted the $\star$-products in the quadratic terms of the Lagrangian since upon integration one $\star$-product disappears in the action and thus they have no influence on the equations of motion. They could have some topological effect through the boundary terms, but such effects are not relevant here.

In the following, we consider, as customarily (see, e.g., \cite{Bahns2}), the Hamiltonian \eqref{total_Hamilt_int_p}, without worrying about the corresponding Lagrangian.
The point of interest here is the way $\star$-products in the
interaction Hamiltonian modify the time evolution. Using again the
integral representation \eqref{int rep}, we have:
\begin{align}
i \frac{\partial}{\partial t}   H^\star_{int}(t)  =& \left[  H^\star_{int}(t), H_0 \right] \notag\\
=&\int  d^3 x d^3y\left[ \frac{\lambda}{3!}  (\phi\star\phi\star\phi)(t,\vect y), \mathcal H_0(t,\vect x)\right] \notag\\
=&\frac{\lambda}{2\cdot 3!}\int  d^3x\, d^3y \int d^4a_1 \,d^4a_2\, d^4a_3 \frac{1}{\pi^4\det \theta} \mathcal K( y; a_1,a_2,a_3) \times \notag\\
&\left[\phi(a_1)\phi(a_2)\phi(a_3), (\partial_0\phi(x)
)^2+(\partial_i \phi(x))^2 + m^2\phi^2(x)\right] \,.
\end{align}
As the time coordinates $a_i^0$ are integration variables, the
commutator will again depend on $\Delta(a_i-x)$. Explicitly, we have:
\begin{align}
\frac{1}{2}[\phi(a_i),  (\partial_0\phi(x))^2 &+(\partial_j \phi(x))^2 + m^2\phi^2(x)]=\notag \\
&= \partial_0 \Delta(a_i-x)  \partial_0 \phi(x)+ \partial_j
\Delta(a_i-x)  \partial_j \phi(x) + m^2\Delta(a_i-x)  \phi(x)\,.
\end{align}
Since this expression is integrated over $x$, it can be further simplified by partially integrating in the
spatial directions and remembering that the fields satisfy the free-field equations:
\begin{align}
\int d^3x\, &\left( \partial_0 \Delta(a_i-x)  \partial_0 \phi(x)+ \partial_j \Delta(a_i-x)  \partial_j \phi(x) + m^2\Delta(a_i-x)  \phi(x) \right) \notag \\
&= \int d^3x\, \left( \partial_0 \Delta(a_i-x)  \partial_0 \phi(x) -\Delta(a_i-x) (\partial_j^2-m^2)\phi(x) \right) \notag \\
&= \int d^3x\, \left( \partial_0 \Delta(a_i-x)  \partial_0 \phi(x)
-\Delta(a_i-x) \partial_0^2\phi(x) \right) \,.
\end{align}
Thus, we can write the evolution as
\begin{align}
 \left[  H^\star_{int}(t), H_0 \right] =&  \frac{\lambda}{3!}\int  d^3x\, d^3y \int d^4a_1 \,d^4a_2\, d^4a_3 \frac{1}{\pi^4\det \theta} \mathcal K( y; a_1,a_2,a_3) \times \notag\\
&\Bigl[ \left(\partial_0 \Delta(a_1-x)  \partial_0 \phi(x)-\Delta(a_1-x) \partial_0^2\phi(x)\right) \phi(a_2)\phi(a_3)\notag\\
&+\phi(a_1)\left(\partial_0 \Delta(a_2-x)  \partial_0 \phi(x)-\Delta(a_2-x) \partial_0^2\phi(x)\right) \phi(a_3)\notag\\
&+\phi(a_1)\phi(a_2)\left(\partial_0 \Delta(a_3-x)  \partial_0 \phi(x)-\Delta(a_3-x) \partial_0^2\phi(x)\right) \Bigr] \,.
\end{align}

The terms proportional to $\Delta(a_i-x)$ vanish only when $a_i^0$ coincides with $t$ and thus we get contributions from all other times including the distant future, as shown in the previous subsection. Thus the evolution of the interaction Hamiltonian at the time $t$ is influenced by field configurations in its future, signalling the lack of causality. One does not obtain anymore the usual tautology, meaning that the interaction Hamiltonian has a non-trivial time-dependence, therefore the energy of the interacting system is not conserved. This result is connected to the lack of integrability condition \eqref{int}. It is also the reason why the equation of motion of the interacting quantum field in this setting differ from the ordinary ones \cite{Bahns2}.

The fact that in the Tomonaga-Schwinger formalism the energy-momentum conservation requires the integrability condition was first shown by K. Nishijima \cite{Nishijima_en_mom_cons}.

\subsection{Comments}

The proposal of quantization of noncommutative field theory in Ref.
\cite{DFR2}, re-stated in \cite{Bahns2}, is essentially based on the
Hamiltonians of interaction defined as
\be\label{Hamiltonian_unit} H_{int}(t)=\int_{x_0=t} d^3 x
(\phi\star\phi\star\ldots\star\phi)(x)\,\ee
and, assuming that the $S$-matrix exists, the time-ordering is taken
with respect to the overall times of the Hamiltonians of
interactions, (and not with respect to the ``times of the fields'') as:
\be\label{time_ord_unit} S=\sum_{n=0}^\infty \frac{(-i)^n}{n!}
\int_{-\infty}^{+\infty} dt_1 \ldots\int_{-\infty}^{+\infty} dt_n
\theta(t_1 - t_2)\ldots\theta(t_{n-1} - t_n){H}_{int}(t_1)\ldots
{H}_{int}(t_n)\,. \ee
The resulting $S$-matrix is indeed unitary, as has been shown time
and again \cite{Bahns2} (see also \cite{Grosse03,LS02}). It is well known, on
the other hand, that the path integral approach, in which the
time-ordering is inherently taken with respect to the times of the
individual fields composing the Hamiltonian of interaction, shows
violation of unitarity \cite{GM,Fuji}.

The choice of time-ordering adopted in Refs. \cite{DFR2,Bahns1} is
the most natural in Hamiltonian formulation and has indeed been used
in theories with higher-derivative couplings, in which case the
presence of the time derivatives in the Hamiltonian of interaction
leads to its dependence on the normal to the hypersurfaces $\sigma$
appearing in the Tomonaga-Schwinger equation. However, the
time-ordered product of derivatives of field operators also depends
on the choice of the hypersurfaces $\sigma$, and thus it is not
Lorentz-covariant. These two effects conspire to render the $S$-matrix
independent of the choice of the family of space-like hypersurfaces
$\sigma$, as it should be, and it can be shown that:
\bea
S=1+\sum_{n=1}^\infty\frac{(-i)^n}{n!}\int_{-\infty}^\infty\ldots
\int_{-\infty}^\infty\, d^4x_1\ldots d^4 x_n\, T[\mathcal
H_{int}(x_1)\ldots \mathcal H_{int}(x_n)]\\\label{Matthews_th_H}
=1+\sum_{n=1}^\infty\frac{(-i)^n}{n!}\int_{-\infty}^\infty\ldots
\int_{-\infty}^\infty\, d^4x_1\ldots d^4 x_n\, T^\star[\mathcal
L_{int}(x_1)\ldots \mathcal
L_{int}(x_n)]\,,\label{Matthews_th_L}\eea
where $T^\star$ is the covariant modification of the usual
time-ordering:
\be T^\star\left[\frac{\partial\phi(x)}{\partial
x^\mu}\frac{\partial\phi(y)}{\partial y^\nu}\right]=
\frac{\partial}{\partial x^\mu}\frac{\partial}{\partial
y^\nu}T[\phi(x)\phi(y)]\neq T\left[\frac{\partial\phi(x)}{\partial
x^\mu}\frac{\partial\phi(y)}{\partial y^\nu}\right]\,.\ee
Remark that for a theory with higher-derivative interactions,
\be\label{H-L}
\mathcal H_{int}(x)\neq -\mathcal L_{int}(x),
\ee
as we have commented also in the previous subsection.

Eq. \eqref{Matthews_th_L} represents the so-called Matthews' theorem,
first noted by P.T. Matthews \cite{Matthews} on the concrete example
of the interacting nucleon-scalar meson fields with vector coupling
and elegantly generalized by K. Nishijima
\cite{Nishijima_higher_deriv} (see also \cite{Koba, Rohrlich}). It
should be pointed out that the path integral quantization naturally
leads to covariant time ordering of Lagrangian densities, as in
\eqref{Matthews_th_L}.

The $\star$-product in the Hamiltonian of interaction
\eqref{Hamiltonian_unit} can indeed be seen as producing
higher-derivative couplings, and the question arises whether
Matthews' theorem is applicable. The answer is negative, since the
proof of the theorem is essentially based on the uniqueness of the
solution of Tomonaga-Schwinger equation
\cite{Nishijima_higher_deriv}, which can be achieved in a theory
with a finite number of time derivatives in the interaction term,
but not when the number of such derivatives is infinite as in the
$\star$-product in the case of a noncommutative time.


The solution of the Tomonaga-Schwinger equation being not unique,
various time evolutions can occur,
leading to different predictions. The choice \eqref{time_ord_unit}
is just one pragmatic possibility. Such a evolution is no doubt
conceivable, however, the nonconservation of energy and momentum for
such a system is an undesirable feature, which can not be cured by the time-ordering procedure.

From a technical point of view, it is noteworthy that the  $T$ time-ordering prescription of \cite{Bahns2}, while preserving unitarity, violates the positive-energy prescription: ``negative-energy'' particle in the sense of Feynman can propagate in the forward time direction \cite{Fuji}, thus forbidding the Wick rotation to Euclidean theory (upon the rotation, the contour of integration cuts through the poles).

\subsection{String theoretical insights for nonconservation of energy}

The energy nonconservation and the unitarity problem appear to be interconnected in a natural way when we invoke open string theory in a background field \cite{SW}. Strong enough background electric field on the $D$-brane can lead to instability of the vacuum by Schwinger pair production. Noncommutative time corresponds to an electric field background in string theory, but in this case the noncommutative field theory as the low-energy limit is not attainable \cite{SW,GM}, since precisely in that limit massive open strings are not irrelevant. The violation of energy-momentum conservation can be easily explained in this context, since the field theory does not take into consideration the whole Hilbert space of states and there are interactions with exchange of energy among the states considered and those ignored in the noncommutative field theoretical calculation. Thus, embedded in this larger context, the system described by the quantum field theory with noncommuting time is naturally nonconservative.

In \cite{LAG} the issue of unitarity violation was carefully examined within noncommutative quantum field theory and interpreted in terms of ghostly tachyonic particles which are produced in the scattering, hence unifying the string theoretical and the quantum field theoretical no-go arguments against noncommuting time.

\section{Noncommutative time and Yang-Feldman-K\"all\'en approach}
\setcounter{equation}{0}

In the study of nonlocal theories, the Heisenberg-picture approach by Yang, Feldman and K\"all\'en \cite{Yang-Feldman, Kallen} has customarily been used \cite{Kris,Hayashi,Marnelius_action, Marnelius_Smatrix}. In this approach the interaction picture is avoided entirely and the elements of the $S$-matrix are calculated directly in the Heisenberg picture with the help of the $n$-point correlation functions (Green's functions). In theories with time-space noncommutativity, the Yang-Feldman-K\"all\'en approach has been considered for example in \cite{Doescher, BahnsThesis, Zahn2}.

The study of nonlocal field theories in the Heisenberg picture, especially of the nonlocal Kristensen-M{\o}ller model \cite{Kris},  was a popular subject in the 1950s. A general treatment with utmost care to all details has been performed in the beginning of the 1970s by R. Marnelius \cite{Marnelius_action, Marnelius_Smatrix}, with the definite conclusion that a unitary $S$-matrix can not be defined for relativistic nonlocal field theories in general. An action principle is given in \cite{Marnelius_action} with quantum variations, since the usual c-number variations lead to nonconserved quantities in such non-canonical systems. The new action principle leads to new forms of the integral conserved quantities, while requiring as consistency conditions the stationarity of the action and the uniqueness of the integral conserved charges. The existence of a unitary $S$-matrix would automatically follow from the consistency conditions. However, these conditions are not satisfied in relativistic nonlocal theories and a unitary $S$-matrix can not therefore be defined.

The conclusions of Marnelius hold also for relativistic time-space noncommutative quantum field theories, for which the Poincar\'e-invariant form factor is the kernel of the star-product. For the completeness of our argumentation, we shall present very briefly the main assumptions and conclusions of \cite{Marnelius_Smatrix}, for the ``noncommutative'' Kristensen-M{\o}ller model \cite{Kris}, described by the Lagrangian:
\begin{align}
&\mathcal{L}(x) = \frac{1}{2}(\partial_\mu \phi(x) \partial^\mu \phi(x)-\mu^2\phi^2(x)) + i \frac{1}{2}\bar{\psi}(x)\slashed \partial \psi(x)-m\bar{\psi}(x)\psi(x)+ \mathcal{L}_{int}(x)\,, \notag\\
&\mathcal{L}_{int}(x) = -g \int  d^4a_1 d^4a_2d^4a_3 K(x;a_1,a_2,a_3)  i \bar{\psi}(a_1) \gamma_5 \phi(a_2)\psi(a_3) \,.
\end{align}
From translational invariance it follows that the kernel can be written in terms of the coordinate differences
\begin{equation}
 \mathcal{L}_{int}(x) = -g \int  d^4\xi d^4\eta K(\xi, \eta)  i \bar{\psi}(x+\eta) \gamma_5 \phi(x)\psi(x+\xi)\,,
\end{equation}
where we have introduced the notation $\xi = a_3-a_2$, $\eta= a_1-a_2$ and $x=a_2$ to match the conventions of \cite{Marnelius_Smatrix}. The equations of motion are:
\begin{align}\label{YFeqs}
( \partial_\mu \partial^\mu+ \mu^2 )\phi(x) = -g &\int d^4\xi d^4\eta\; K(\xi, \eta) \bar{\psi}(x+\eta)i\gamma_5\psi(x+\xi)=-g\rho(x)\,,\notag\\
i(\slashed \partial -m)\psi(x) = g&\int  d^4\xi d^4\eta\; K(\xi, \eta) i\gamma_5 \phi(x-\eta)\psi(x-\eta+\xi)=gf(x)\,,
\end{align}
with the obvious meaning for the $\rho$ and $f$.
The solutions are written in the integral form, according to the Yang-Feldman-K\"all\'en procedure, in terms of either the $in$- or $out$-fields, as
\begin{align}\label{YFK_solution}\nonumber
&\phi(x) = \phi_{in}(x)-g\int d^4y \Delta_R(x-y) \rho(x) \quad;\quad\phi(x) = \phi_{out}(x)-g\int d^4y \Delta_A(x-y) \rho(x)\,,\nonumber\\
&\psi(x) = \psi_{in}(x)+g\int d^4y S_R(x-y) f(x)\quad;\quad\psi(x) = \psi_{out}(x)+g\int d^4y S_A(x-y) f(x)\,.
\end{align}
Here $\Delta_R$, $\Delta_A$, $S_R$ and $S_A$ are the usual retarded and advanced Green's functions for bosonic and fermionic fields. The asymptotic $in$ or $out$ fields defined at $t\rightarrow \mp \infty$ satisfy free-field equations. The lack of causality is manifest in \eqref{YFK_solution}, since the behaviour of an interacting quantum field at a given space-time point is determined by its whole past {\it and future} history. There seems to be a conflict between the equations of motion and the boundary conditions, but the above formal solutions are assumed.

The solutions of \eqref{YFeqs} can be expressed iteratively as a series expansion in the {\it in}-fields or alternatively in the {\it out}-fields,
\begin{align}\label{fieldexp1}
&\phi(x) = \phi_{in/out} (x) + \sum_{n=1}^{\infty}g^n\phi^{(n)}(x;in/out)\,, \notag\\
&\psi(x) = \psi_{in/out} (x) + \sum_{n=1}^{\infty}g^n\psi^{(n)}(x;in/out)\,,
\end{align}
where the $\phi^{(n)}(x;in/out)$ and $\psi^{(n)}(x;in/out)$ are functionals of the $in$ or $out$ fields, respectively. For the theory to be consistent, the same interacting fields $\phi(x)$ and $\psi(x)$ should be obtained by using either expansion. This turns out not to be the case for the model under consideration.

Unlike the local case, the interacting fields here are not canonical variables (which can be seen immediately from the fact that $\phi(x)$ and $\psi(x)$ do not commute) and the Schwinger action principle can not be used in its original form. Allowing also $q$-number variations \cite{Marnelius_action} and requiring to retain the equations of motion \eqref{YFeqs}, in \cite{Marnelius_Smatrix} it was shown that the infinitesimal generators of $c$-number variations, $F_{c}(t)$, should be modified as
\begin{equation}\label{modifiedF}
F_{c}(t) \rightarrow F_{t_0}(t) = F_c(t) -\frac{g}{2} \int_{t_0}^t d^4x\ \delta_0 A(x) \,,
\end{equation}
where  $\delta_0 A(x)$ in the Kristensen-M{\o}ller model is given by
\begin{equation}
\delta_0 A(x) = \int d^4\eta \, d^4\xi F(\xi , \eta) \left([\delta_0 \phi(x), \bar{\psi}(x+\eta)]i\gamma_5\psi(x+\xi)-\bar{\psi}(x+\eta)]i\gamma_5[\delta_0 \phi(x), \psi(x+\xi)] \right) \,.
\end{equation}
The $in-$ and $out$ -representations of generators are then given by the limits:
\begin{align}\label{inandout}
&F_0(t;in)=\lim_{t_0\rightarrow -\infty } F_{t_0}(t)\,,  \notag \\
&F_0(t;out) = \lim_{t_0\rightarrow +\infty } F_{t_0}(t)    \,.
\end{align}
Using \eqref{modifiedF} and \eqref{inandout}, we get the difference of the $in$ and $out$ representations of generators
\begin{equation}\label{genDiffs}
F(t;out) - F(t;in) = \frac{g}{2} \int_{-\infty}^{+\infty}d^4x\ \delta_0 A(x) \,.
\end{equation}
Specifically, for the momentum generators (for details on how the momentum generators are obtained, we refer the reader to the original paper \cite{Marnelius_Smatrix}) this difference is
\begin{align}
P^\nu(t;out) -&P^\nu(t,in) = \;\frac{g}{2}\int d^4x \, d^4\eta \, d^4\xi K(\xi , \eta)\notag\\
& \times \left( \bar{\psi}(x+\eta)i\gamma_5 [\partial^\nu\phi(x), \psi(x+\xi)]-  [\partial^\nu\phi(x), \bar{\psi}(x+\eta)] i\gamma_5 \psi(x+\xi) \right) \,.
\end{align}
When the fields are expanded as in \eqref{fieldexp1} it was shown in \cite{Marnelius_Smatrix} that in fourth order in $g$ the difference of $P^\nu$'s does not vanish but is given by
\begin{align}\label{PDiff}
P&^\nu(t;out) -P^\nu(t,in) = -ig^4\int \prod_{i=1}^4 d^4x_i d^4\xi_i d^4\eta_i K(\xi_1,\eta_1)K(\xi_2,\eta_2)K(\xi_3,\eta_3)K(\xi_4,\eta_4) \notag\\
 & \times\Theta(x_1+\xi_1-(x_2+\eta_2))\Theta(x_2-x_4)\Theta(x_4+\eta_4-(x_3+\xi_3))\Theta(x_3-x_1)\partial_4^\nu \Delta(x_4-x_2)\Delta(x_1-x_3)  \notag\\
 &\quad \times\bar{\psi}_{in}(x_1+\eta_1)\bar{\psi}_{in}(x_3+\eta_3)S(x_3+\xi_3-(x_4+\eta_4))i\gamma_5 \psi_{in}(x_4+\xi_4)\notag\\
 &\quad\quad \times  S(x_1+\xi_1-(x_2+\eta_2))i\gamma_5 \psi(x_2+\xi_2)+h.c. + O(g^5) \,,
\end{align}
where $\Theta(x)$ is the Heaviside step function.
The nonvanishing of \eqref{PDiff} will result in the expansions \eqref{fieldexp1} in terms of $in$- and $out$-fields giving different expressions for the interacting quantum fields $\phi(x)$ and $\psi(x)$. This follows from the requirement that the momentum generators perform the transformations that they are assumed to perform
\begin{align}
&[P^\nu,\phi(x)] = -i\partial^\nu \phi(x)\,, \notag\\
&[P^\nu,\psi(x)] = -i\partial^\nu \psi(x)\,,
\end{align}
irrespective of the representation. If the fields in the two representations coincided, we would have, for example,
\begin{equation}
[P^\nu(t;out)-P^\nu(t;in), \psi(x)] = 0 \,.
\end{equation}
But since the difference of the momentum generators \eqref{PDiff} is proportional to the fields $\psi_{in}$ and $\bar{\psi}_{in}$ at different times, the commutator is nonzero in fourth order in $g$. It follows that the interacting fields derived from the $in$-fields can not be the same as those derived from the $out$-fields.

The nonuniqueness of solutions, in turn, leads to the nonstationarity of the action for both sets of solutions. This is because the variation of the total action turns out to be equal to the difference of the generators in the $in$ and $out$ representations \label{genDiff} $$\int\delta \left(d^4x \mathcal{L}(x)\right)= F(t;out) - F(t;in) = \frac{g}{2} \int_{-\infty}^{+\infty}\delta_0 A(x).$$

As a consequence, there does not exist a unitary $S$-operator that would relate the $in$- and $out$- fields by a similarity transformation:
\begin{align}\label{Smatrix}
\phi_{out}(x) &= S^{-1}\phi_{in}(x)S\,, \notag\\
\psi_{out}(x) &= S^{-1}\psi_{in}(x) S\,.
\end{align}
Rather, it was shown in \cite{Marnelius_Smatrix} (see also \cite{Imamura}) that from
\begin{equation}
\psi_{out}(x) = \psi_{in}(x) -g\int d^4y S(x-y)f(y)
\end{equation}
a direct calculation in fourth order $g$ gives
\begin{equation}
S^\dagger\psi_{in}(x)S = \psi_{out}(x)+g^4(\cdots) \neq \psi_{out}(x) \,,
\end{equation}
i.e. there is no unitary $S$-operator satisfying \eqref{Smatrix} in this picture.

In conclusion, if we consider that quantum fields satisfy the equations of motion in the Yang-Feldman-K\"all\'en approach, the infinitesimal generators will be modified and the field expressions in the $in$ or $out$ representations will not coincide. This discrepancy further leads to the nonexistence of a unitary $S$-matrix in the Heisenberg picture.

\subsection{Comments}

The lack of unitarity in the Yang-Feldman-K\"all\'en approach for time-space noncommutative quantum field theories has never been spelled out so far (see, however, Ref. \cite{Schroer}, for a critical review). Direct calculations have only been done up to the second order in the coupling constant.
However, doubts about the asymptotic completeness have been expressed in \cite{Bahns2}.

Also, it was noticed in \cite{BahnsThesis} that the interacting field obtained in the Hamiltonian and Yang-Feldman-K\"all\'en formalisms are different. Besides other inconsistencies in the theory, this aspect has a simple explanation, since the Yang-Feldman-K\"all\'en approach makes use of the Lagrangean formalism and, as mentioned in subsection \ref{energy_noncons}, the Lagrangian and the Hamiltonian of interaction are not the opposite of each other for higher derivative couplings in general. Simply, the two interacting fields which are compared are obtained for two different theories.

\section{Aspects of path integral quantization with noncommutative time}
\setcounter{equation}{0}

In commutative space-time the existence of a path integral is deeply connected with the time ordering of Heisenberg picture operators. When we use Schwinger's action principle in the path integral quantization (see the monographs \cite{Fujikawa,Chaichian-Nelipa,Chaichian-Demichev}), we usually assume that the vacuum-to-vacuum transition amplitude, from infinitely distant past to infinitely distant future, in the presence of a localized source, is:
\begin{equation}\label{v-to-v}
\langle 0,+\infty |0,- \infty\rangle _J = \int {\mathcal D}\phi \ exp\ \left[i\int
d^4 x {\mathcal L}_J\right]\,,
\end{equation}
where, for instance,
\be \mathcal L_J(x) = \frac{1}{2}\partial^\mu\phi(x)\partial_\mu\phi(x)+
 \frac{1}{2}
m^2\phi^2(x)-\frac{\lambda}{3!}\phi^3 (x)+\phi(x)J(x).\ee
Green's functions and hence all physics are given through Schwinger's action principle by functional derivatives of the path integral as, for example,
\begin{equation}\label{Greens}
\langle 0, +\infty |T^\star\hat{\phi} (x) \hat{\phi} (y)|0,-\infty \rangle = \frac{\delta}{i\delta J(x)} \frac{\delta}{i\delta J(y)} \langle0, +\infty |0,- \infty\rangle _J\bigg|_{J=0}\,,
\end{equation}
where $|0,\pm \infty\rangle _J$ are the asymptotic vacuum states at the times
$t=\pm\infty$ in the presence of a source function $J(x)$ localized in space-time. In the construction of Green's functions, space-time is sliced with Heisenberg picture states and this choosing of the ``path'' automatically selects $T^\star$ as the time ordering, i.e. time ordering with respect to the {\it times of the fields}.

Eq. \eqref{v-to-v} connects the operator formalism in the Heisenberg picture with the path integral formalism. The existence of this connection for noncommutative field theories with nonlocal time is not \emph{a priori} obvious. In ordinary quantum field theory, for a scalar field for example, the connection between the operator formulation and the path integral approach is established by making use of the path integral for the quantum mechanical oscillator, with its creation and annihilation operators. This is known to be a solution of the Schr\"odinger equation, with the well-known meaning of the transition amplitude. What is sought for is the path integral representation for the transition amplitude
\be
\langle \phi_f,t_f|\phi_i,t_i\rangle=\langle \phi_f|e^{-iH(t_f-t_i)}|\phi_i\rangle\,,
\ee
expressed in the Heisenberg picture.
The space is discretized, the phase-space path integral for a countable number of quantum mechanical harmonic oscillators is used, the continuum limit is taken at the end and the result is
\be\label{pathint_phase_space}
\langle \phi_f,t_f|\phi_i,t_i\rangle=\int {\mathcal D}\phi(x) \frac{{\mathcal D}\pi(x)}{2\pi}\exp{\left[i\int_{t_i}^{t_f}d^4x \left(\pi(x)\partial_t\phi(x)-\mathcal H(x)\right)\right]}\,.
\ee
This is the fundamental formula connecting the Heisenberg picture transition amplitudes -- as long as they are consistent and well-defined -- to the path integral representation in phase-space. For the case of Hamiltonians quadratic in $\pi(x)$, the Gaussian integrals over  $\pi(x)$ can be performed and the Lorentz-covariant form of the path integral is obtained:
\be\label{pathint_covariant}
\langle \phi_f,t_f|\phi_i,t_i\rangle=\int {\mathcal D}\phi(x) \exp{\left[i\int_{t_i}^{t_f}d^4x {\mathcal L}(x)\right]}\,.
\ee
Upon taking the asymptotic times $t=\pm\infty$, only the vacuum states in \eqref{pathint_covariant} survive, since the other strictly positive energy states give rapidly oscillating contributions which add up to zero. In the end, formula \eqref{v-to-v} is obtained in the presence of a local source term.
However, it is not obvious in general that formula \eqref{pathint_phase_space} can be brought to a path integral over the fields only, \eqref{pathint_covariant}, if the Hamiltonian is not quadratic in the canonically conjugated momenta.

The whole procedure of connecting the operator formalism to path integrals has to be repeated for relativistic noncommutative field theories. Obviously, this is impossible, since the very Heisenberg picture is inconsistent \cite{Marnelius_Smatrix} and conceptually it makes little sense to recast it in a different form \footnote{Remark that the Lorentz-covariant path integral formulation exists for relativistic quantum field theories with derivative couplings and the connection between the Hamiltonian and Lagrangian pictures is given precisely by Matthew's theorem \eqref{Matthews_th_L}.}. Moreover, according to \cite{Marnelius_action}, Schwinger's action principle does not hold with $c$-number variations, but a new action principle allowing for operator-valued variations has to be introduced, which makes implausible the connection with the path integral which does not involve any operators at all.

It is customary to start from operator formalism and obtain the path integral formulation. However, there are well-established methods of going in the opposite direction, i.e. to recover a canonical structure from a path integral. The prescription for this is due to J. D. Bjorken, K. Johnson and F. E. Low \cite{BJL} and it has been explored in a convincing manner in the context of field theories with noncommutative time in \cite{Fuji}. As is to be expected, a whole set of inconsistencies is recovered, but no canonical structure.
The path integral formulation naturally reproduces the unitarity problem of the Heisenberg picture \cite{Fuji}. The advantage is that the results are much easier to come about by path integrals than in the Yang-Feldman-K\"all\'en theory.

Although the path integral formulation is intrinsically related to the canonical quantization, the approach of \cite{Fuji} starting from the path integral expression has been extremely useful for bringing up awareness of the conceptual, not only technical, intricacies of noncommutative time in quantum field theory, especially regarding other problems than the already much discussed unitarity issue.

\section{Conclusion}

The study of relativistic time-space noncommutative quantum field theories is strongly connected to the physically motivated commutation relations of coordinates \cite{DFR1,DFR2}. We have re-examined these field theories in various pictures and quantization procedures: the interaction picture Hamiltonian formulation (Tomonaga-Schwinger equation), the Yang-Feldman-K\"all\'en approach in  the Heisenberg picture and the path integral method. Conceptual inconsistencies of various kinds are present in all these approaches: i) in {\it interaction picture}, the main inconsistency is the non-uniqueness of the solution of the equation of motion and the energy-momentum nonconservation; ii) in {\it Heisenberg picture}, the failure of the unitarity of the $S$-matrix is manifest in the fourth order in the coupling constant (for a Kristensen-M{\o}ller-type of noncommutative model, but it is clearly not improved for other interactions) \cite{Marnelius_Smatrix}; iii) in the {\it path integral formulation}, the path integral generating functional itself is meaningless in the absence of a sound Heisenberg picture, and this principal inconsistency surfaces as the impossibility to derive by Bjorken-Johnson-Low prescription a canonical structure \cite{Fuji}.

We can conclude that there is no panacea for (relativistic) time-space noncommutative quantum field theories -- any attempt to cure one problem has not been able to render the whole theory consistent. The dominant feature which is the source of all the inconsistencies is the violation of causality, manifested as infinite nonlocality in time. But infinite nonlocality, even in space, is responsible for the standing problem of space-space and also light-like noncommutative theories -- the UV/IR mixing problem \cite{MSV}, which hinders renormalizability. Moreover, the $\theta\to 0$ limit in any quantum correction is not a smooth limit to ordinary QFT, but rather leads to singularities.

It is thus the infinite nonlocality that one has to restrict in a consistent manner and it is to be expected that a success in this direction would make the $\theta\to 0$ limit also smooth. One possible approach is to introduce the noncommutativity of space-time in a dynamical manner. It is also conceivable that Lorentz invariance should be given up at high energies, if the gravity effects are really to be probed by the noncommutativity of space-time, while in the low-energy limit Lorentz symmetry would be recovered. Such a proposal, for a Very Special Relativity, has been recently made in \cite{VSR} and has been connected to light-like noncommutativity in \cite{NCVSR}. Finding a mechanism for such a ``symmetry breaking on the reverse'' may provide an answer to the nonlocality problems in noncommutative quantum field theory.

\section*{Acknowledgements}

We are much grateful to Masud Chaichian for useful discussions and suggestions and for his never-ceasing interest in this work. We are indebted to Jos\'e Gracia-Bond\'ia, Robert Marnelius, Peter Pre\v{s}najder and Shahin Sheikh-Jabbari for important comments on the manuscript. AT cherishes the memory of the late Professor Kazuhiko Nishijima and recalls many instructive discussions over the years, regarding the subject of this paper. The support of
the Academy of Finland under the Projects No. 136539 and 140886 is
gratefully acknowledged.

\vskip1cm



\end{document}